# High Curie Temperature Ferromagnetism and High Hole Mobility in Tensile Strained Mn-doped SiGe Thin Films

*Huanming Wang, Sen Sun, Jiating Lu, Jiayin Xu, Xiaowei Lv, Yong Peng, Xi Zhang, Yuan Wang, Gang Xiang\**

Prof. G. Xiang, Prof X. Zhang, H. Wang, Dr. J. Lu, J. Xu,
College of Physics, Sichuan University, Chengdu 610064, China
Email: gxiang@scu.edu.cn

Prof. Y. Wang, Dr. S. Sun
Key Laboratory of Radiation Physics and Technology, Ministry of Education,
Institute of Nuclear Science and Technology, Sichuan University, Chengdu 610064, China

Prof. Y. Peng, X. Lv
Key Laboratory for Magnetism and Magnetic Materials of Ministry of Education,
Lanzhou University, Lanzhou 73000, China

**Abstract**

Diluted magnetic semiconductors (DMSs) based on group-IV materials are desirable for spintronic devices compatible with current silicon technology. In this work, amorphous Mn-doped SiGe thin films were first fabricated on Ge substrates by radio frequency magnetron sputtering and then crystallized by rapid thermal annealing (RTA). After the RTA, the samples became ferromagnetic (FM) semiconductors, in which the Curie temperature increased with increasing Mn doping concentration and reached 280 K with 5% Mn concentration. The data suggest that the ferromagnetism came from the hole-mediated process and was enhanced by the tensile strain in the SiGe crystals. Meanwhile, the Hall effect measurement up to 33 T to eliminate the influence of anomalous Hall effect (AHE) reveals that the hole mobility of the annealed samples was greatly enhanced and the maximal value was ~1000 $cm^2 \cdot V^{-1} \cdot s^{-1}$, owing to the tensile strain-induced band structure modulation. The Mn-doped SiGe thin films with high Curie temperature ferromagnetism and high hole mobility may provide a promising platform for semiconductor spintronics.

Dilute magnetic semiconductors (DMSs) prepared by doping transition metal (TM) such as Cr, Fe and Mn into nonmagnetic semiconductors have generated much interest due to their potential applications in the new classes of spintronic devices. An ideal DMS should offer various characteristics including high Curie temperature (ideally, above room temperature) carrier-mediated ferromagnetism and high carrier mobility,[1,2] which are essential for spin polarized current generation and transport in real spintronic devices. In order to synthesize ideal DMSs, substantial work has been carried out on different materials such as III-V,[3–6] II-VI,[7] group-IV[8,9] and I–II–V[10] and oxide[11,12] semiconductors. Achievements have been made on increasing Curie temperature of the DMSs. For instance, Curie temperature of the model material (Ga, Mn)As that exhibits hole-mediated ferromagnetism has been increased to 200 K by nanostructure engineering[6] and transition-metal doped ZnO shows defect-related ferromagnetism above room temperature.[12] However, carrier mobility in the DMSs is relatively low, owing to the defects introduced during the synthesis process. For instance, experimental studies show that the hole mobility in (Ga,Mn)As is only up to 10 $cm^2 \cdot V^{-1} \cdot s^{-1}$ and the electron mobility in (In,Fe)As is several tens of $cm^2 \cdot V^{-1} \cdot s^{-1}$.[13] Another interesting fact is that, although extensive transport studies of the DMSs have been performed,[1,3–6,9,10,12–14] the studies on the carrier mobility in the DMSs were rare, probably because anomalous Hall effect (AHE) in the DMSs makes the accurate estimation of hole mobility notoriously difficult.[14] To the best of our knowledge, very few DMSs have been reported that can exhibit both high Curie temperature ferromagnetism and high carrier mobility simultaneously.

Band structure modification by strain engineering is usually used for enhancing hole mobility in traditional semiconductors.[15–17] Among group-IV semiconductors, Si and Ge are completely miscible, which makes SiGe alloy become one of the ideal materials for band structure modification by controlling Si and Ge concentrations. Since tensile strain can separate heavy holes and light holes on the valence band top and increase hole mobility in SiGe alloys[17] and enlarging lattice constant also benefits the ferromagnetic (FM) order in Mn-doped group-IV semiconductors,[18] we speculate that tensile-strained Mn-doped SiGe thin films on Ge substrates may be promising for high-performance spintronic materials. However, although there were reports on the studies of Mn-doped Si or Ge,[8,9,19–21] no investigation on the tensile-strained Mn-doped SiGe thin films has been reported.

In this work, we report the results on the fabrication and structural, magnetic and electrical properties of tensile-strained $Si_{0.25}Ge_{0.75}$:$Mn_x$ films with different nominal Mn concentration $x$. Magnetism and high-field magneto-transport measurements show that we obtained a high Curie temperature (280 K) and a high hole mobility (~1000 $cm^2 \cdot V^{-1} \cdot s^{-1}$) in the FM $Si_{0.25}Ge_{0.75}$:$Mn_x$ films. Further experimental characterizations and theoretical analysis were then performed to explore the correlation between the tensile strain, dopants, ferromagnetism and hole mobility in the annealed samples.

A series of $Si_{0.25}Ge_{0.75}$:$Mn_x$ samples were first prepared by radio frequency (r.f.) magnetron sputtering. For simplicity, the three $Si_{0.25}Ge_{0.75}$:$Mn_x$ samples in this work are named as M1 ($x$=0.01), M2 ($x$=0.025) and M3 ($x$=0.05), respectively. **Figure 1**a shows a cross-sectional transmission electron microscope (TEM) image of a typical as-grown sample M2. All the as-grown samples show similar microstructures. The gray-scale high angle annular dark-field (HAADF) image and energy dispersive spectroscopy (EDS) image are shown in Figure S1 (Supporting Information). It indicates that the top layer and the buffer layer of sample M2 were homogeneous. The High-resolution transmission electron microscope (HRTEM) image of the M2 top layer shown in Figure 1b shows an amorphous phase with an amorphous ring in the fast Fourier transform (FFT) image. It was found that Ohmic contact couldn't be made on the as-grown amorphous samples.

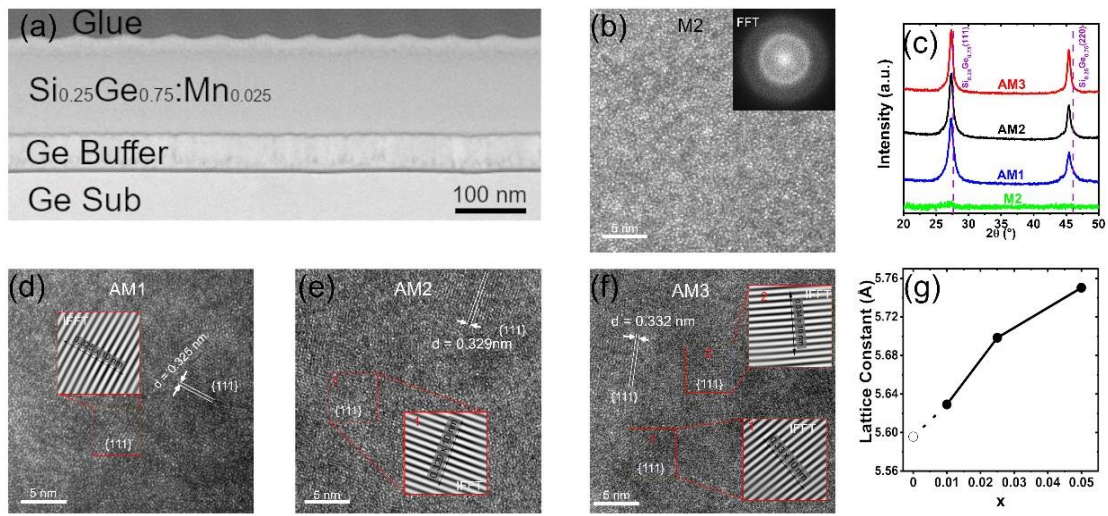

**Figure 1.** (**a**) Cross-sectional transmission electron micrograph of sample M2. (**b**) The HRTEM image of sample M2. The inset shows the FFT image. (**c**) GIXRD patterns of AM1, AM2, AM3 and M2. HRTEM images for AM1(**d**), AM2(**e**) and AM3(**f**). The insets show the IFFT image and the interplanar distance. (**g**) Mn concentration vs. lattice constant.

To improve the crystalline quality and activate the dopants in the samples, a careful nonequilibrium rapid thermal annealing (RTA) treatment was performed. For convenience, we name the annealed M1, M2 and M3 samples as AM1, AM2 and AM3, respectively. Figure 1c shows the grazing incidence X-ray diffraction (GIXRD) patterns for AM1, AM2, AM3 and M2. Obviously, the intensity of the peaks of annealed samples is much larger than the as-grown sample and the full width at half maximum (FWHM) is also much narrower. These indicate that the RTA improved the crystallinity quality of samples. Careful search for secondary phases such as $Ge_8Mn_{11}$ or $Ge_3Mn_5$ was performed. However, even in the logarithmical scale, the GIXRD patterns show only the diamond structure of SiGe but no other crystalline structures (In contrast, the typical secondary phase in GIXRD pattern of heavily doped AM4 with *x*=0.08 is shown in Figure S2a in Supporting Information). The SiGe grain sizes in AM1, AM2 and AM3 calculated by Scherrer formula are 7.4 nm, 8.6 nm and 10.1 nm, respectively. Following Vegard's law, the lattice constant of the unstrained $Si_{0.25}Ge_{0.75}$ is 5.595Å. All the GIXRD peak positions of the annealed samples had a left-shift with respect to the unstrained values, indicating that there exists tensile strain in the samples. HRTEM investigation also reveals the formation of SiGe crystals after RTA (Figure 1d-f). Like in the GIXRD measurement, no secondary phases could be found in HRTEM measurement. With an FFT filtering and following inverse FFT (IFFT) operation, the lattice constant could be acquired (Figure 1d-f inset IFFT pattern). The relationship between Mn concentration and lattice constant is displayed in Figure 1g. Since the atomic radius of Mn (1.61 Å) is larger than those of Ge (1.25 Å) and Si (1.11

Å), the fact that the lattice constant increases with increasing Mn concentration indicates that the Mn atoms were incorporated into the $Si_{0.25}Ge_{0.75}$ matrix.[22] Considering that the lattice constant of Ge substrate is larger than that of $Si_{0.25}Ge_{0.75}$, the tensile strain in $Si_{0.25}Ge_{0.75}$ matrix was probably caused by both Mn incorporation and Ge substrate.

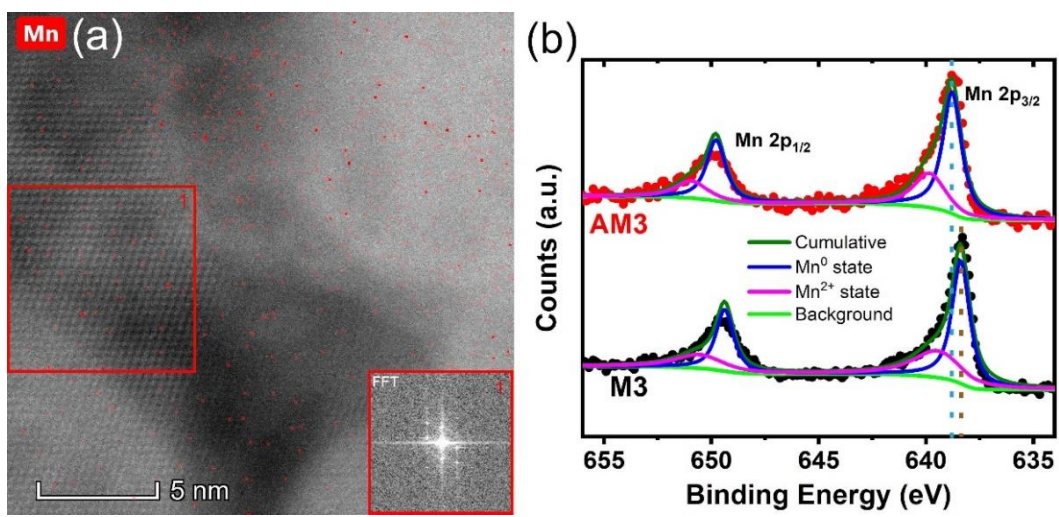

**Figure 2**. (**a**) Cross-sectional ACTEM image of AM2. The inset shows the FFT image. (**b**) XPS spectra results of AM3 and M3.

The spherical aberration corrected transmission electron microscope (ACTEM) was used to detect the distribution of Mn atoms in the annealed samples. Mn atoms were found in both the crystalline phase and amorphous phase in the annealed samples and the Mn concentration was around a few percent. All the annealed samples show similar results. AM2 is taken as an example here and the result is presented in **Figure 2**a. Considering the low Mn equilibrium solubility ($<10^{-7}$ *at.*%) in the Ge and Si lattice,[23] the incorporation of such more Mn atoms in the diamond-structured SiGe lattice (red square) should be caused by the nonequilibrium RTA process, which is similar to the fact that the nonequilibrium growth process of GaMnAs allows a large amount of Mn atoms to be incorporated into the host GaAs lattice far above its solubility.[3]

To figure out the Mn chemical states before and after RTA, we investigated the chemical binding states with X-ray photoelectron spectroscopy (XPS). As an example, in Figure 2b, the XPS spectra of M3 and AM3 fitted with the Lorentzian-Gaussian function are shown. The sample pairs of M1 and AM1 and M2 and AM2 show similar results. These spectra have been charge corrected with respect to the adventitious C $1s$ peak, where the asymmetrical peaks result from the multi-electron process.[24] In the as-grown sample, the $Mn2p_{3/2}$ peak at 638.4 eV mainly corresponds to zero valence state of Mn, indicating that Mn in the amorphous phase mainly has a zero valence state (blue line). Also, the $Mn2p_{3/2}$ peak appeared at 639.4 eV is due to the contribution of $Mn^{2+}$ state.[25,26] After RTA, some zero valence state Mn atoms were converted to the divalent state ones, which resulted in the shift of the cumulative peak to a higher binding energy region by 0.4 eV. Since a Mn atom shows a zero valance state in an interstitial site and shows a divalent state in a substitutional site,[27] the increasing amount of divalent state Mn atoms indicates that the RTA induced substitutional doping of Mn atoms in the SiGe matrix.[28] In addition, comparing the result of sample AM3 and M3, no high atomic valence of Mn was detected (In contrast, $Mn^{4+}$ could be detected in AM4 in Figure S2b in Supplementary Material) by XPS and no manganese metal was detected by HRTEM in the annealed samples. So, the RTA process not only crystallized the as-grown SiGe samples but also induced the substitutional doping and thermal activation of Mn atoms in the SiGe matrix. As a result, Mn dopants provided both the holes and local magnetic moments in the annealed samples.

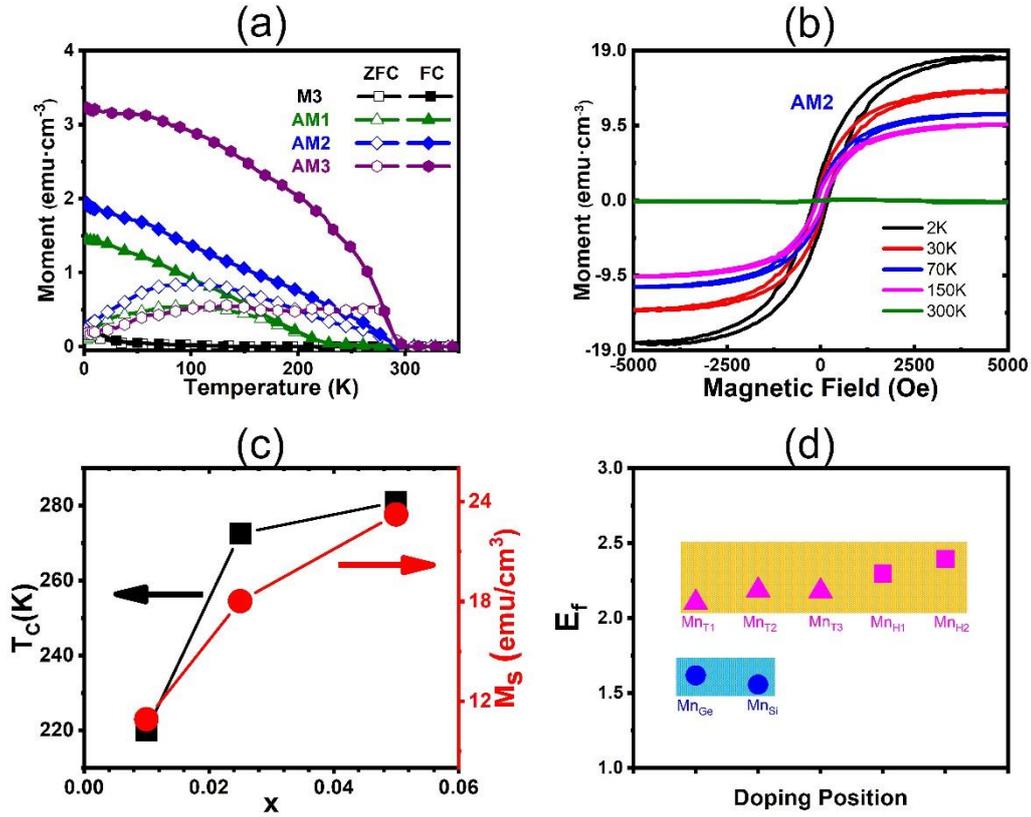

**Figure 3** (**a**) Temperature dependence of the magnetization of AM1, AM2, AM3 and M3. (**b**) M-H curve for AM2 at different temperatures. (**c**) Curie temperature ($T_C$, left) and saturation magnetic moment ($M_S$, right) of AM1, AM2 and AM3. (**d**) The formation energies of doping a Mn atom at different positions in the $Si_{0.25}Ge_{0.75}$ lattice. $Mn_{Ge}$ ($Mn_{Si}$) represents the Mn atom replacing one Ge (Si) atom. $Mn_T$ ($Mn_H$) represents the Mn atom occupying an interstitial site with tetrahedral (hexagonal) symmetry in the diamond lattice. According to the symmetry, there are two possible situations for $Mn_H$ and three possible situations for $Mn_T$.

The magnetic properties of the annealed samples were measured by a superconducting quantum interference device (SQUID). **Figure 3**a shows the temperature dependence of zero-field cooling (ZFC) and field cooling (FC) magnetization for AM1, AM2 and AM3. The results for M3 are also shown for comparison. Under the ZFC condition, the samples were cooled down to low temperature (2 K in our measurement) without an applied magnetic field and then heated to high temperature with a fixed magnetic field (20 Oe in our measurement). The ZFC magnetization of the annealed samples first increases then decreases as the temperature increases. Unlike the ZFC condition, the samples were cooled down and then heated up with an applied magnetic field (20 Oe) in the FC process. The FC magnetization of the annealed samples achieves its maximum at the lowest temperature, then decreases with increasing temperature and drops to zero near the Curie temperature. In contrast, in the control sample of M3, there exists a weak spontaneous magnetization in the whole temperature range both in ZFC and FC conditions, which may be induced by Mn diluted in the minute quantities of nanocrystals formed during the sputtering process. In Figure 3b, the M-H curves of AM2 measured with a perpendicular magnetic field show clear hysteresis loops below 300 K and the coercive fields decrease as the temperature increases, indicating a FM ordering below room temperature in the annealed sample. All the annealed samples show similar results. The Curie temperature $T_C$ and saturation magnetization $M_S$ as the function of the nominal Mn concentration is shown in Figure 3c, where $T_C$ is defined by differentiation of magnetization over temperature in M-T curve and $M_S$ is extracted from M-H curve at 2 K. As the Mn concentration increases, the $T_C$ increases ($T_C$ for AM1, AM2 and AM3 is 220 K, 273 K and 280 K, respectively) and so does the $M_S$ ($M_S$ for AM1, AM2 and AM3 is 11 emu·cm$^{-3}$, 18 emu·cm$^{-3}$ and 23 emu·cm$^{-3}$, respectively). It is worthwhile to note that $Ge_8Mn_{11}$ and $Ge_3Mn_5$, the most commonly found GeMn compounds, both exhibit a $T_C$ at 300 K,[29,30] and $Ge_8Mn_{11}$ shows an antiferromagnetic (AFM) order below 150 K. However, none of these magnetic features were detected here. On the contrary, the $T_C$ of our annealed samples was found to increase with Mn doping concentrations. Also, only a diamond-structured SiGe phase was detected and no $Ge_8Mn_{11}$ or $Ge_3Mn_5$ clusters were observed in the structural measurements of the annealed samples. Thus, the ferromagnetism in our annealed samples should come from Mn dopants dispersed in the $Si_{0.25}Ge_{0.75}$ matrix.

To understand the high $Tc$ in the annealed samples, we should point out that the ferromagnetism in Mn-doped $Si_{0.25}Ge_{0.75}$ crystals comes from hole-mediated process, in which holes would itinerate around and align the local magnetic moments of the Mn ions in one direction through the exchange coupling.[1,9] We performed first-principle calculations by using the Vienna *ab initio* simulation package (VASP) within the generalized gradient approximation (GGA). A 2×2×2 supercell with 64 atoms (48 Ge atoms and 16 Si atoms) arranged in the lowest energy configuration was chosen to simulate the $Si_{0.25}Ge_{0.75}$ lattice. As shown in Figure 3d, the calculation of the formation energies of a Mn atom doped at different sites shows that the Mn atom prefers to substitute a Ge or Si lattice atom than locate in an interstitial site, either one tetrahedrally coordinated or one with hexagonal symmetry in the diamond lattice.[31]

The energetically favorable substitutional Mn dopants provide both holes and local spins in the $Si_{0.25}Ge_{0.75}$ lattice, which is similar to the case of substitutional Mn-doped Ge. [9,32] Previous studies have shown that substitutional Mn-doped Ge is a Ruderman-Kittel-Kasuya-Yosida (RKKY)-like FM semiconductor which exhibits FM state at the lowest energy configuration,[32] and substitutional Mn-doped Si also tends to be FM but with lower energy differences between AFM and FM states[18]. So, with the same doping concentration, the Curie temperature of $Si_{0.25}Ge_{0.75}:Mn_x$ will probably be higher than that of Mn-doped Si. Furthermore, enlarging lattice constant (tensile strain) can decrease the p-d mixing and benefit the FM order in Mn-doped group-IV semiconductors.[18] Thus, the energetically favorable substitutional Mn doping and tensile strain in $Si_{0.25}Ge_{0.75}$ gives rise to such as high $Tc$.

As shown in **Figure 4**a, the temperature dependence of the resistivity for AM2 under zero filed shows that the sample is on the insulating side of the metal-insulator transition which is typical for doped semiconductors.[33] All the annealed samples show similar behavior. At low temperatures, the resistivity increases exponentially with decreasing temperature and becomes infinite as temperature trends to zero. At high temperature, the resistivity shows a trend to saturate. Below 25 K, the relationship between resistivity and temperature could be fitted by an exponential law ($T^{-a}$) with $a \approx 0.275$, close to the value 0.25 for variable range hopping.

It is notoriously difficult to accurately determine carrier concentration and mobility in the DMSs because of the presence of AHE. Generally, the Hall resistance $R_{xy}$ is expressed as $R_{xy} = R_0 \cdot B \cdot d^{-1} + R_S \cdot M \cdot d^{-1}$, where $R_0$ and $R_S$ represent the ordinary and anomalous Hall coefficients, respectively, $d$ the film thickness and $M$ the sample magnetization. Since the ordinary Hall effect term ($R_0 \cdot B \cdot d^{-1}$) keeps increasing as the magnetic field $B$ increases but the AHE term ($R_S M \cdot d^{-1}$) reaches saturation as $B$ is sufficiently high, the slope of curve $R_{xy} \sim B$ can be extracted at the high field (33 T in our measurement) and the ordinary Hall effect can be separated from the AHE to determine the actual carrier concentration and mobility

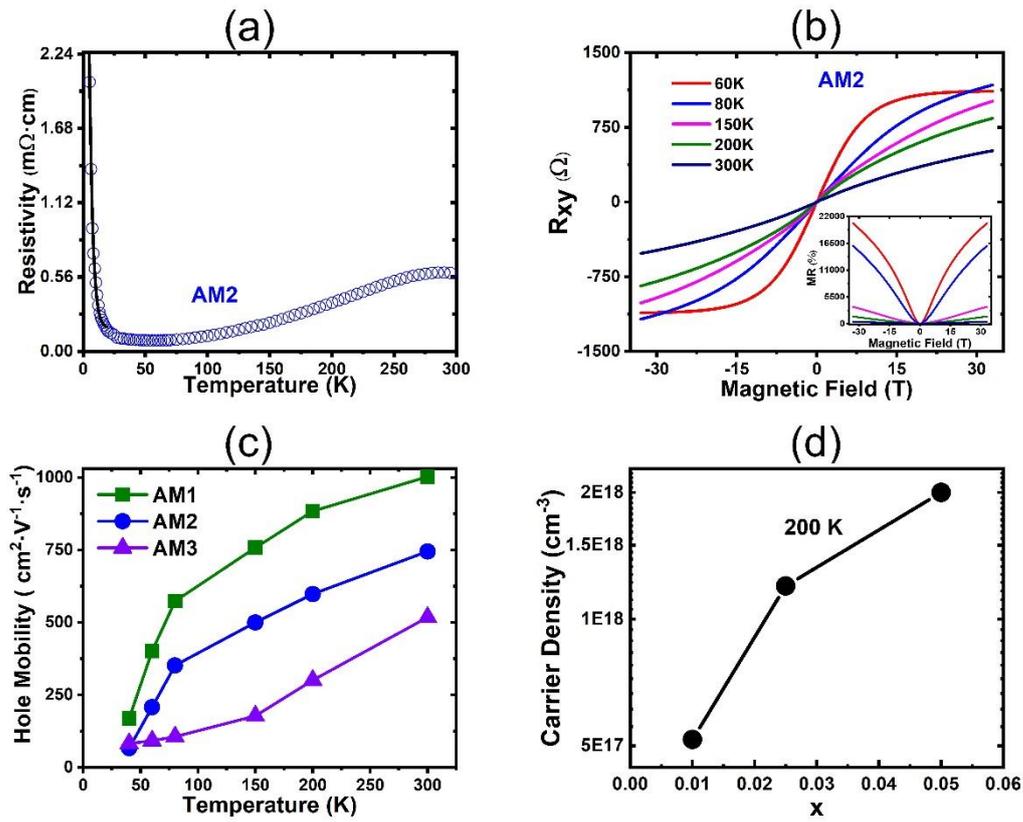

**Figure 4.** (**a**) Temperature dependence of resistivity under a zero magnetic field for AM2 (blue open circles). The solid line corresponds to variable range hopping. (**b**) Magnetic field dependence of Hall resistance and MR (inset) for AM2 at different temperatures. (**c**) Temperature dependence of the hole mobility for AM1, AM2 and AM3. (**d**) The hole density as the function of the Mn doping concentration at 200 K.

Figure 4b shows the magnetic field dependence of the Hall resistance measured at different temperatures for AM2. The clear AHE signals at the temperatures below 300 K manifest the FM ordering in the annealed samples below the $T_C$, which is another key characteristic of DMS ferromagnet and consistent with the SQUID measurement results. The positive slopes at high fields indicate that the majority carriers are holes. The magnetoresistance (MR) for AM2 is shown in the inset of Figure 4b, which could be up to 20000% at 33T at 60 K. The large MR can be attributed to the geometrically enhanced MR.[34,35] In zero magnetic field, the electrical current flowing through the crystalline SiGe grains (Figure 2a) with low resistivity is parallel to the electric field, while in the high magnetic field, the current is no longer parallel to the electric field owing to the Lorentz force and is deflected from the grains to amorphous phase with high resistivity. The transition from the low resistivity at zero magnetic field to high resistivity at high magnetic field results in the observation of the large MR.[34,35] Figure 4c shows the temperature dependence of the obtained hole mobility ($\mu$) for AM1, AM2 and AM3. Interestingly, the hole mobility of the strained samples can be enhanced up to 1003 cm$^2 \cdot$V$^{-1} \cdot$s$^{-1}$ at 300 K. It is known the hole mobility in the unstrained Si$_{0.25}$Ge$_{0.75}$ samples is ~300 cm$^2 \cdot$V$^{-1} \cdot$s$^{-1}$ at 300 K,[36,37] smaller than ~ 450 cm$^2 \cdot$V$^{-1} \cdot$s$^{-1}$ in Si and ~1900 cm$^2 \cdot$V$^{-1} \cdot$s$^{-1}$ in Ge owing to the alloy scattering. However, the tensile strain results in the removal of the valence band degeneracy (and hence the reduction of the effective hole mass) and the reduction of intervalley scattering[17,38] and finally increase the hole mobility in Si$_{0.25}$Ge$_{0.75}$:Mn$_x$ samples despite of the scatterings. As a result, we obtained a maximal hole mobility of 1003 cm$^2\cdot$V$^{-1}\cdot$s$^{-1}$, ~2 orders of magnitude higher than those of other DMSs including (Ga, Mn)As.

In all the annealed samples, the hole mobility decreases with the decreasing temperature, which obviously cannot be fitted by acoustic phonon scattering ($\mu \sim T^{-1.5}$). For AM3, $\mu$ can be fitted by a power law ($\mu \sim T^{1.8}$, Figure S3 in Supporting Information) which is close to the ionized impurity scattering relationship ($\mu \sim T^{1.5}$), but for AM1 and AM2, the power law fittings show large errors. Since the scatterings from both ionized Mn substitutional atoms and Mn interstitials exist in the doped samples, this indicates that as the Mn doping concentration increases, more Mn substitutional atoms are activated and the ionized scattering becomes more evident, which is consistent with the relationship between the hole density and Mn doping concentration in the AM samples shown in Figure 4d.

Si$_{0.25}$Ge$_{0.75}$:Mn$_x$ samples were fabricated by using r.f. magnetron sputtering and nonequilibrium RTA process. Structural characterizations reveal the polycrystalline structure and the tensile strain inside the annealed samples. The annealed samples exhibited a Curie temperature up to 280 K and the ferromagnetism in the annealed samples were found to originate from the substitutional Mn doping in the Si$_{0.25}$Ge$_{0.75}$ lattice and enhanced by tensile strain. The annealed samples achieved a hole mobility up to 1003 cm$^2\cdot$V$^{-1}\cdot$s$^{-1}$ estimated by the Hall effect at a high magnetic field of 33 T, which was enhanced by the tensile strain induced band engineering and ~ 2 orders of magnitude higher than those in other DMSs. The discovery of the high Curie temperature and high hole mobility in the $Si_{0.25}Ge_{0.75}:Mn_x$ samples may provide a promising platform for the semiconductor spintronics.

**Experimental Section**

Sample Preparation and Characterization: The $Si_{0.25}Ge_{0.75}:Mn_x$ films were fabricated on intrinsic Ge (001) substrate by using r. f. magnetron sputtering through co-sputtering Ge (99.9999%) target, Si (99.9999%) target and Mn (99.99%) target. The detailed process is as follows. The Ge wafers were cleaned with deionized water, ethanol and acetone, followed by 2.5% HF etching. After chemically clean, the Ge wafers were loaded into the chamber of the magnetron sputtering machine. The base pressure prior to deposition was about $1 \times 10^{-5}$ Pa. A 50 nm Ge layer was firstly deposited on the substrate as a buffer sublayer. Subsequently, a 150 nm $Si_{0.25}Ge_{0.75}:Mn_x$ top layer was deposited on the Ge buffer sublayer. During the deposition, the substrate was kept at 250 °C and the pressure was kept at 0.3 Pa $Ar_2$ atmosphere. The Mn composition ($x$) was controlled by changing the sputtering power of the Mn target. After deposition, samples were annealed at 800℃ for 30s by using RTA furnace (NBD-HR1200-110IT) in 95%$Ar_2$/5%$H_2$ atmosphere. Films without sublayer were deposited on $Al_2O_3$ substrate for detecting the composition by Energy-dispersive X-ray spectroscopy (EDX) in SEM measurement. XPS (Thermo Scientific Escalab250Xi) measurement was carried out to study the chemical state of our samples. To avoid the influence of surface contaminates, a 50nm thick Ar ion thinning was applied before the XPS measurement. GIXRD (Empyrean Panalytical diffractometer Cu $K$-$\alpha$ with $\lambda = 0.154$ nm) and TEM (FEI Tecnai F30) were applied to detect the structural properties of the samples. The cross-section samples for the TEM measurement were prepared by a focus ion beam (FIB) milling procedure in a Helios (Tescan LYRA 3 XMU). The distribution of Mn elements inside the SiGe matrix was detected by using ACTEM, (FEI Titan G2 300). Magnetic properties were measured by using SQUID magnetometer (Quantum Design MPMS XL-7). The electrical transport properties were investigated by using a Hall bar-shaped device (In this case, length = 0.44 cm, width = 0.04 cm) which is fabricated by using standard photolithography and wet etching process. Indium is used to obtain Ohmic contact on our annealed samples. However, we could not obtain any Ohmic contact on the as-grown samples owing to the low carrier density in the amorphous phase. During the electrical transport measurement, we used a water-cooling magnet (WM-5) at the High Magnetic Field Laboratory at Heifei, Chinese Academy of Sciences. The current was 10 μA, modulated at a frequency of 13.7 Hz by using a Keithley 6221. The transverse and longitudinal voltages were measured by an SR830 Lock-In Amplifier.

Simulation: Our first-principles calculations were performed using the plane-wave technique as implemented in the Vienna *ab initio* simulation package (VASP)[39]. The exchange-correlation potentials were approximated by a local density approximation (GGA)[40]. A 350 eV cutoff for the plane-wave basis set was adopted in all simulations. The convergence criteria were set as $10^{-5}$ eV in energy and 0.01 eV·Å$^{-1}$ in force. In the self-consistent field and total energy calculations, a Monkhorst–Pack *k*-point sampling of 4×4×4 was used. All atomic positions were optimized by minimizing total energy and atomic force. First of all, an unit-cell composed of 2 Si atoms and 6 Ge atoms was composed, which corresponded to the $Si_{0.25}Ge_{0.75}$ lattice, and the lowest energy configuration of the unit-cell was explored and found. Next, we expanded the lowest energy configuration into a 2×2×2 supercell (64 atoms). Finally, a Mn atom was doped into the supercell. According to the symmetry, there were 7 possible doping situations: two situations ($Mn_{Si}$ and $Mn_{Ge}$) for substitutional Mn doping on Si and Ge site, three situations ($Mn_{T1}$, $Mn_{T2}$ and $Mn_{T3}$) for interstitial Mn doping on the tetrahedral position, two situations ($Mn_{H1}$ and $Mn_{H2}$) for interstitial Mn doping on the hexagonal position of the diamond lattice. Figure S3 in Supporting Information shows doping different situations, in which the unit cell was used to demonstrate the atom positions for simplicity.

**Supporting Information**

Supporting Information is available from the Wiley Online Library or from the author.

**Acknowledgements**

We thank the support from the National Key R&D Plan (Grant No. 2017YFB0405702) and the National Natural Science Foundation of China (Grant No. 51671137). We thank the High Magnetic Field Laboratory of Chinese Acadamy of Sciences for the help of the high magnetic field transport measurements of our samples. We thank Prof. Jianhua Zhao, Dr. Hailong Wang and Prof. Mingliang Tian for assisting in data collection and analysis.

Received: ()
Revised: ()
Published online: ()

# Supporting Information

**High Curie Temperature Ferromagnetism and High Hole Mobility in Tensile Strained Mn-doped SiGe Thin Films**

*Huanming Wang, Sen Sun, Jiating Lu, Jiayin Xu, Xiaowei Lv, Yong Peng, Xi Zhang, Yuan Wang, Gang Xiang\**

The HADDF/EDS mapping image of sample M2 is shown in Figure S1. The distribution of elements(Ge, Si, and Mn) in M2 top layer is homogeneous. Typical second phase measurement results of sample AM4 could be seen in Figure S2. Clear $Ge_3Mn_5$ second phase could be seen in GIXRD pattern (Figure S2a). In Figure S2b, the $Mn2p_{3/2}$ peak appeared at 641.7 eV is considered to the contribution of $Mn^{4+}$ state which is induced by the $Ge_3Mn_5$. In Figure S3, we give the seven possible doping situations of the Mn dopant in $Si_{16}Ge_{48}$. In Figure S4, the $\mu$-$T$ curve fitting result of sample AM3 is shown, which indicated that the ionized scattering is the dominant scattering mechanism.

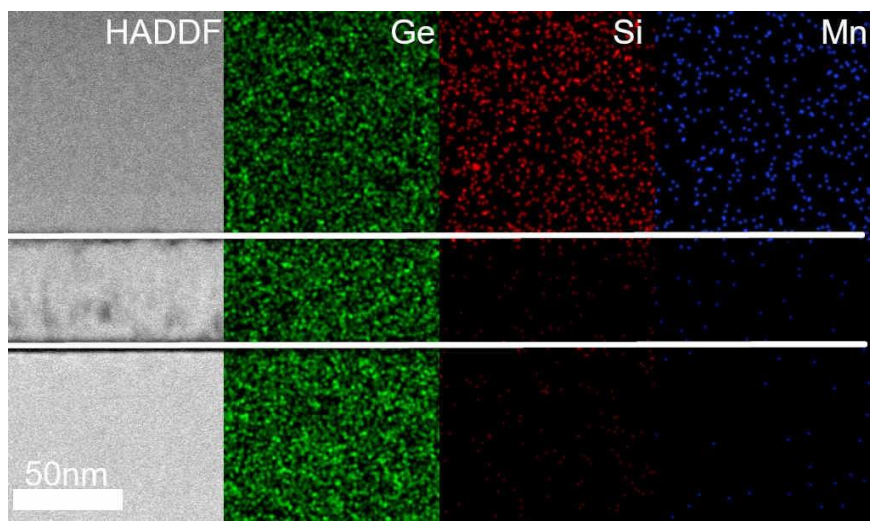

**Figure S1.** HADDF/EDS mapping image of sample M2.

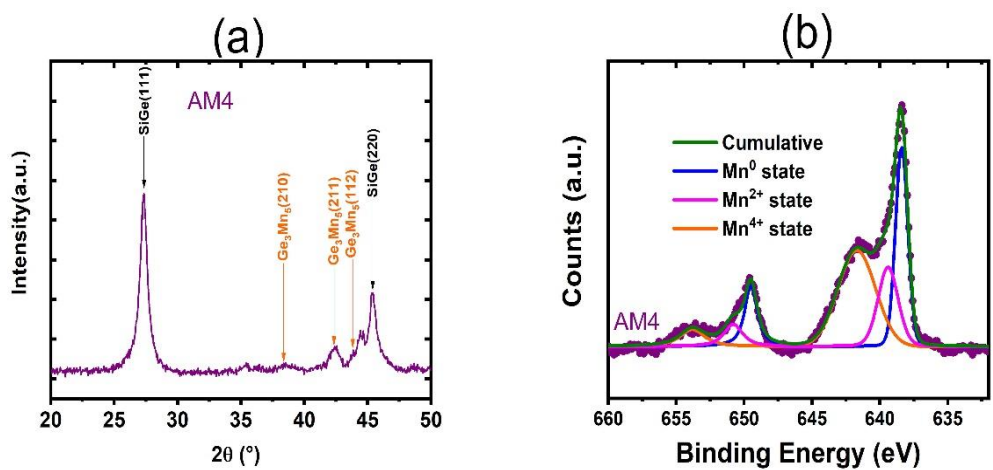

**Figure S2.** (a) XRD pattern for sample AM4. (b) XPS spectra for sample AM4.

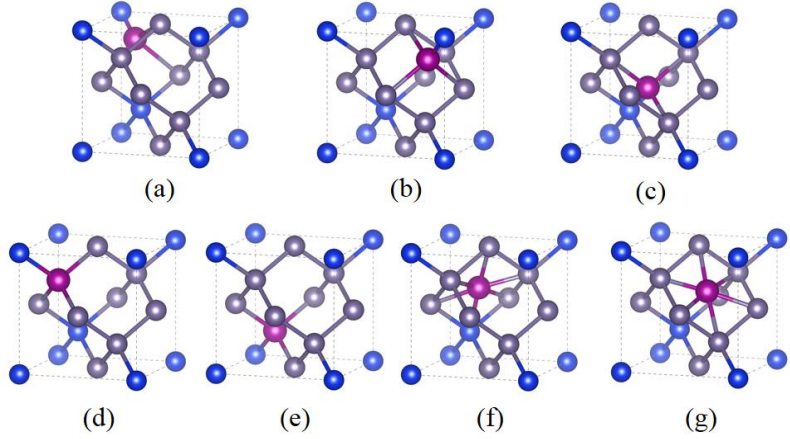

**Figure S3.** For a clearer illustration, we used the $Si_2Ge_6$ unit-cell as a schematic. (a-c) Mn located in three different interstitial tetrahedral positions ($M_T$). (d) Mn replacing Ge ($Mn_{Ge}$). (e) Mn replacing Si ($Mn_{Si}$). (f-g) Mn located in two different interstitial hexagonal positions ($Mn_H$). Blue atoms represent for Si atoms, gray atoms represent for Ge atoms, and purple atoms represent for Mn atom.

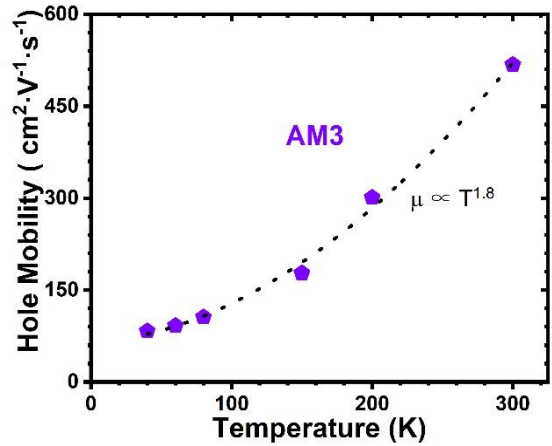

**Figure S4.** $\mu$-$T$ curve fitting result for sample AM3. The dashed line corresponds to the fitting result ($\sim T^{1.8}$).